# ExSampling: a system for the real-time ensemble performance of field-recorded environmental sounds

Atsuya Kobayashi
Keio University
Faculty of Environment and
Information Studies
5322 Endo, Fujisawa, Kanagawa
252-0882 Japan
t16366ak@sfc.keio.ac.jp

Reo Anzai
Keio University
Faculty of Environment and
Information Studies
5322 Endo, Fujisawa, Kanagawa
252-0882 Japan
t16510ra@sfc.keio.ac.jp

Nao Tokui
Keio University
Graduate School of Media and
Governance
5322 Endo, Fujisawa, Kanagawa
252-0882 Japan
tokui@sfc.keio.ac.jp

## ABSTRACT

We propose ExSampling: an integrated system of recording application and Deep Learning environment for a real-time music performance of environmental sounds sampled by field recording. Automated sound mapping to Ableton Live[1] tracks by Deep Learning enables field recording to be applied to real-time performance, and create interactions among sound recorders, composers and performers.

## 1. INTRODUCTION

After Pierre Schaeffer established the concept of music concrete by recording and processing sounds in cities and nature, his experimental methods were applied to today's popular musical expressions. The technique of collecting specific sounds by field recording and deforming and modulating the sounds has spread as a computer music expression technique and has been used by various artists.

On the other hand, when demonstrating computer music performance using these environmental sounds, artists must select sound samples from recorded sounds by listening to them, and process them such as equalizing and map them to music scenes and musical instruments. It takes a long time for artists before starting to use those sounds in their composition. For this reason, the number of specific sound samples that can be used is limited and the real-time nature of the performance is lost.

Even though sounds are constantly generated throughout the world, the time spends on processing and travelling causes the loss of real-time nature when used in live performance afterwards. We pursed to maintain the nature of the field recordings in music performance, by using sound that originated somewhere in the world in real-time.

For this purpose, we have devised a system that enables real-time field recording to be applied to music performances, using Deep Learning techniques. We propose to automate the selection of environmental sound samples.

To the best of our knowledge, this paper is the first to propose the application of machine learning for employing field-recorded sounds in real-time musical expression.

In the previous studies that took a similar approach, [1] proposed a system that automatically mashes up songs by quantitatively evaluating the similarity between songs. In [2], a versatile system made up of modules that detect pitch and beat communicating via OSC (Open Sound Control)[2] enables real-time mash-up of input sounds. Also, recognition of music genre and musical instruments are possible with high accuracy by Deep Learning technology [3]. In our research, we applied these strategies to environmental sound recognition and mapping to MIDI instruments and samplers.

This paper describes the technical implementation and interface design of the system, introduces application cases for performance, and then describes the main contributions and future development of this research.

## 2. METHODOLOGY

System implementation and interface design

## 2.1 SYSTEM OVERVIEW

The system consists of the following three modules. The first module is a Web-based sound recorder for collecting environmental sounds. The recording function is implemented using recorder.js[3], a wrapper library for the Web Audio API[4], and the sound recorded by the device's microphone can be immediately sent to the performer's computer. By publishing the recorder application to the Internet through network port

---

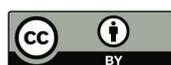

[1] https://www.ableton.com/en/live/

[2] http://opensoundcontrol.org/
[3] https://github.com/mattdiamond/Recorderjs
[4] https://www.w3.org/TR/webaudio/

forwarding using ngrok[5], anyone can participate in the recording of the performance.

**Fig 1. System Overview**

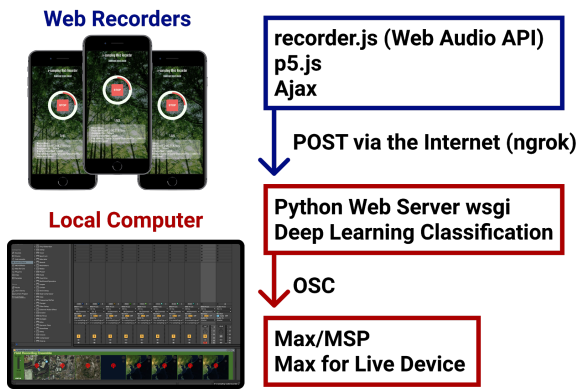

The second module is a Web API server and a Deep Learning execution environment written in Python[6]. This module also functions as a back end for providing a sound recorder. After cutting the part of the received sound file of less than 20db, the module converts it to a spectral image and inputs it to the Deep Learning model to estimate the type of sound. The sound file path, the estimated sound type, and detected sound pitch and location information are sent to Max / MSP[7] (Max for Live[8]) by OSC.

The third module, the Max / MSP patch program, links tracks in the DAW to their sample sounds based on the transmitted sound files and classification results. The buffered sound file is played to the note-in from the mapped MIDI track.

## 2.2 INTERFACE DESIGN

**Fig 2. User Interfaces of Recorder and Max Device**

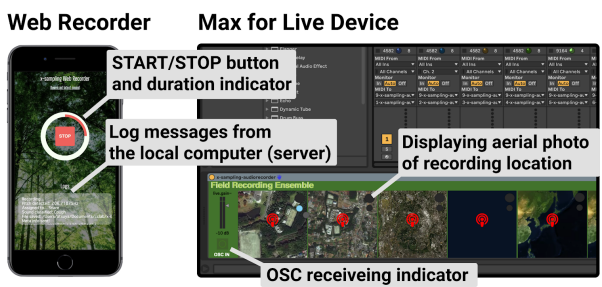

The user interface of the recorder is implemented in p5.js[9] and is designed to make the operation of the field recording easy to understand. Since the log message from the local Web server displays the recognized class of the sound, recording participants can figure out how the sound will be used in the performance. The Max for Live device has a map-type user interface that allows the performer to check where the sounds were recorded. We implemented this interface with Web Geolocation API[10] and MapBox API[11].

## 2.3 SOUND CLASSIFICATION

For the classification of environmental sounds, we adopted MobileNetV2 [4], a lightweight convolutional neural network (CNN) architecture, for low-latency real-time inference on the local laptop computer. To use this CNN model, the sound data were pre-processed using the Python Library Librosa[12]. The recorded environmental sounds are cut into 5-second increments, each of which is converted into a spectrogram. The model reads them as 3 channel images matrix and makes an inference of the class.

**Fig 3. Classification model architecture**

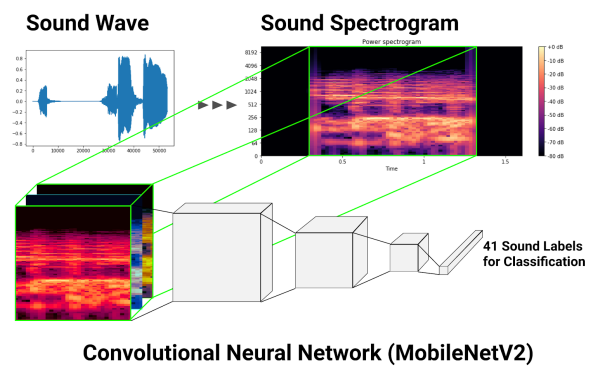

We trained the model using sound data sets with annotated 41 classes [5] that were used and published in the sound recognition competition at Kaggle[13]. We implemented whole training and inference workflow based on open source TensorFlow[14] sample source codes [6].

**Fig 4. Mappings to MIDI Tracks**

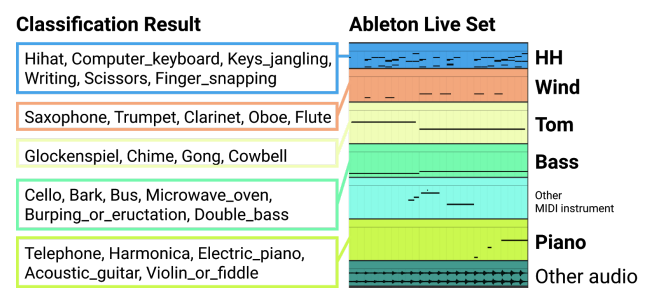

---

[5] https://ngrok.com/
[6] https://www.python.org/
[7] https://cycling74.com/
[8] https://www.ableton.com/en/live/max-for-live/
[9] https://p5js.org/
[10] https://www.w3.org/TR/geolocation-API/
[11] https://docs.mapbox.com/api/
[12] https://librosa.github.io/librosa/
[13] https://www.kaggle.com/
[14] https://www.tensorflow.org/

**Fig 5. Classification labels and default settings for assigned instruments**

| Classification Label | Instrument | Classification Label | Instrument |
|---|---|---|---|
| Bark | Bass | Acoustic_guitar | Piano |
| Burping_or_eructation | Bass | Electric_piano | Piano |
| Bus | Bass | Harmonica | Piano |
| Cello | Bass | Telephone | Piano |
| Double_bass | Bass | Violin_or_fiddle | Piano |
| Microwave_oven | Bass | Cough | Snare |
| Bass_drum | BD | Fireworks | Snare |
| Drawer_open_or_close | BD | Knock | Snare |
| Fart | BD | Shatter | Snare |
| Gunshot_or_gunfire | BD | Snare_drum | Snare |
| Applause | Chorus | Tambourine | Snare |
| Laughter | Chorus | Tearing | Snare |
| Meow | Chorus | Chime | TT |
| Squeak | Chorus | Cowbell | TT |
| Computer_keyboard | HH | Glockenspiel | TT |
| Finger_snapping | HH | Gong | TT |
| Hihat | HH | Clarinet | Wind |
| Keys_jangling | HH | Flute | Wind |
| Scissors | HH | Oboe | Wind |
| Writing | HH | Saxophone | Wind |
| | | Trumpet | Wind |

Sound mapping patterns among the classification results and MIDI instrument tracks can be set by writing configuration file on Python server directory or changing Max for Live device options.

## 3. PERFORMANCE USE CASE

First of all, after training the sound classification model, we prepare the environment for inference. At the time of performance, a web server is set up on the local computer to provide a web-based recorder over the Internet. Using the activation script we prepared, a QR code is issued to access the recorder, so that the recording participants can access the recorder on the Internet via the QR code.

When participants record various environmental sounds with their recorders, the sound data and location information is automatically sent to the performer's local computer, and as soon as it is received, deep learning inferences about the class are made. Instrument mapping based on the classification results is performed on Max, and the samples assigned to MIDI tracks are switched in real-time. When a new sample is received and the sound is switched, the performer will be notified by the Max device and can immediately use the new sampled sound for expression. In this way, the recording participants can contribute to the performance.

The screen recorded video of our sample performance demonstration is available at https://cclab.sfc.keio.ac.jp/paper/exsampling/.

### 3.1 PERFORMER

The performer can assign a categorized sound sample to a premade MIDI track. The MIDI track plays the sample in the specified MIDI note pitch, relative to the original pitch of the sample. We used Librosa to detect the original dominant pitch of each sample. Percussive sounds mapped to hi-hat, tom, snare, etc. are played back in the original pitch, while melodic sounds such as piano, wind, etc. are played back in the same pitch as MIDI notes.

If the performer changes the length of the MIDI note, the duration of the sample being played will also change. The performer can use environmental sounds for both the melody and the beat and mix with computer sound source or their instrument and ensemble with it.

A performer creates MIDI tracks for environmental sound samples, in addition to the regular tracks in Ableton Live. Then creates a MIDI note clip beforehand or create one during the performance to play the ambient sound registered by Max. When a new environmental sound is received, the Max device notifies by indicating location information, and the performer creates a new expression with that.

### 3.2 RECORDING PARTICIPANTS

The recording participants will contribute to the performance with a web recorder, but at the same time, they can watch the performance through online video streaming services and get classification results of their recordings on web recorder, so that they can see how their environment sound material is applied to the musical expression, and feel a sense of contribution to the performance.

Ex-sampling field recording activities, which involve multiple people, also provide an opportunity to consider the soundscape through the rare experience of turning the recorded sounds into music ourselves. Not only performers and composers but also everyone has an opportunity to think about what kind of expression would be interesting if we could immediately turn the sounds of our surroundings into music.

## 4. DISCUSSION AND CONCLUSIONS

The ExSampling system made it possible to apply field-recorded environmental sound to real-time performances. We believe that the method we proposed here encourages more musicians to use field-recorded sounds in their live performances. Based on the method, performers also can sample and assign audio sources to samplers and rhythm machines automatically on stage, and to mix them into pre-recorded tracks they play.

Although we mentioned that it is cumbersome to select sound samples by listening to them one by one, we also understand it is an essential part of the artistic practice of musicians. We don't mean to get rid of this process entirely; rather, we propose another way of selecting sounds. We hope our system works as a pre-selection process: the system filters and classifies samples, from which human musicians can choose. Since our software selects and organizes samples automatically and systematically, it lessens the burden of dealing with samples and allows human musicians to focus on the higher-level musical structure.

We also hope the system produces a sense of "serendipity" in the performance. The automatic sound classification (and misclassification) may lead to unconventional sound mappings, which in turn create interesting musical expression. The performer doesn't know what kind of samples will be provided by the participant in advance, but he or she has loose control over the sound mapping through the classification. This delicate balance between predictability and unpredictability gives the performer amusing tension during the performance.

With the current system, it is not possible to send videos and photos in addition to sound, but this will be possible in a future update. By sending a video of the sound and photos of the place to the performer along with the sound, it is possible for VJs and others who create videos in real-time to apply it to their visual expression and performance. The audience will be able to get a more detailed picture of where the environmental sounds are being recorded. It's an important part of connecting the performance with the context in which the recording participants are present.

Also, there is the disability of performers to select whether or not to use a given environmental sound in their work in real-time. For performers to choose whether or not to incorporate environmental sounds into their work, we need to implement a new interface for the Max for Live device that allows them to listen to the environmental sounds they receive while monitoring their performance.

Further, the classification system needs to improve. The lack of variety and size of datasets limits the diversity of instruments used for tracks. Since the flexibility and utility of automatic sound classification depend on the data used for training the classification model, it is necessary to make a dataset and use it for learning models to classify recorded environmental sounds more suitable for musical expression that the performer wants to do.

It would also be preferable for the performers utilizing this system to be able to train the sound classification model with their datasets. We will tackle these problems and try to improve the usability of the system by integrating all the components into a single software.

The main contribution of this research is that the automation of sample selection in concrete music has made it possible to apply it to real-time performance. By using this system, we proposed a new form of interaction in which many recording participants create a musical performance and culture of live music.